# Reconfigurable ASIC for a Low Level Trigger System in Cherenkov Telescope Cameras

**David Gascon[e], Juan Abel Barrio[a], Oscar Blanch[b], Joan Boix[b], Eric Delagnes[c], Carlos Delgado[d], Lluís Freixas[d], Fabrice Guilloux[c], Ruben López Coto[b,f], Scott Griffiths[b], Gustavo Martínez[d], Oscar Martínez[b], Andreu Sanuy[e] and Luis Ángel Tejedor[a]**

[a] *Universidad Complutense de Madrid (UCM),*
*Ciudad Universitaria, Plaza Ciencias, s/n, 28040, Madrid, Spain*

[b] *Institut de Física d'Altes Energies (IFAE)*
*Edifici CN, Campus UAB, 08193, Bellaterra, Spain*

[c] *IRFU, CEA, Université Paris-Saclay,*
*F-91191 Gif-sur-Yvette, France*

[d] *Centro de Investigaciones Energéticas, Medioambientales y Tecnológicas (CIEMAT)*
*Av. Complutense, 40, 28040, Madrid, Spain*

[e] *Dept. de Física Quàntica i Astrofísica, Institut de Ciències del Cosmos (ICCUB),*
*Universitat de Barcelona (IEEC-UB), Martí Franquès 1, E08028 Barcelona, Spain*

[f] *now at Max-Planck-Institut fur Kernphysik, P.O. Box 103980, D 69029 Heidelberg, Germany*

*E-mail*: dgascon@fqa.ub.edu

ABSTRACT: A versatile and reconfigurable ASIC is presented, which implements two different concepts of low level trigger (L0) for Cherenkov telescopes: the Majority trigger (sum of discriminated inputs) and the Sum trigger concept (analogue clipped sum of inputs). Up to 7 input signals can be processed following one or both of the previous trigger concepts. Each differential pair output of the discriminator is also available as a LVDS output. Differential circuitry using local feedback allows the ASIC to achieve high speed (500 MHz) while maintaining good linearity in a 1 Vpp range. Experimental results are presented. A number of prototype camera designs of the Cherenkov Telescope Array (CTA) project will use this ASIC.

# Contents

## 1. Introduction

Cherenkov telescopes are used to detect Very High Energy (VHE, E > 10 GeV) gamma rays after their interaction with the Earth's atmosphere [1], [2], [3] and [4]. By detecting and processing Cherenkov light produced by gamma-ray-induced Extensive Air Showers (EASs) composed of relativistic electrons and positrons, it is possible to characterize the incoming gamma ray in terms of direction and energy. Placing several telescopes (of different sizes) together to observe the same shower in coincidence results in a gamma-ray detector with enhanced sensitivity over a wide energy spectrum.

Cameras based on fast and sensitive photo-sensors, typically photomultiplier tubes (PMTs) detect Cherenkov light. An electronic readout chain coupled to each photo-sensor digitizes the PMT waveform. The readout chain also incorporates an electronic trigger chain that performs a fast signal processing. Rapid signal processing is necessary for disentangling signals produced by incident Cherenkov light from EASs from those induced by background light from the night sky background (NSB).

In this paper, an Application Specific Integrated Circuit (ASIC) designed specifically for camera trigger decision in Cherenkov telescopes is presented. The ASIC can be configured to perform two different analogue triggering schemes, which we call majority and sum triggering. It also provides an interface for digital trigger systems.

There are a number of motives for implementing multiple trigger schemes in the same chip. First, the majority trigger scheme is a simple and robust system, traditionally implemented in Cherenkov telescopes. In contrast, the sum trigger involves a novel concept (described in Section 2.1), first implemented in the MAGIC telescopes [5], which aims to reduce Cherenkov telescopes' energy threshold as much as possible. Second, it will be very useful for a telescope to have the ability to choose between the two systems depending on different observing conditions. Observing conditions may change as the telescopes is pointed to different regions in the sky with varying NSB levels. In addition, different types of telescopes in CTA can use the same ASIC to implement different trigger schemes. Large Size Telescopes (LSTs), which incorporate 23 m diameter reflectors to study the low energy regime, will need to use the sum trigger scheme when observing weak point sources with high-energy spectra that peaks at lower energies. Medium Size Telescopes (MSTs) might choose the majority trigger for its more robust operation.

Finally, the discriminated output of the individual pixels in low-voltage differential signalling (LVDS) format allows telescopes to implement a digital trigger system, in which the discriminated output of all the pixels in the camera can be combined in search of complex patterns. In this way, images from muons or even hadrons could be identified at trigger level.

Section 2 describes a typical multilevel trigger approach used for Cherenkov telescope cameras [6]. Section 3 presents the architecture of a new versatile ASIC for the implementation of the low level trigger. Section 4 includes measurements of functionality and performance of the ASIC. Conclusions are given in Section 5.

## 2. Trigger architecture

### 2.1 Trigger concept

The trigger decision in Cherenkov telescopes is based on the detection of a concentration of signal in the camera both in space and time. The photons coming from a Cherenkov shower will be detected by several close pixels in a time window of a few nanoseconds. Therefore, a telescope trigger, requiring a given number of pixels of the camera with a minimum number of photons in a short time, will reject a sizeable amount of NSB-induced events because NSB photons are not correlated either in space or in time.

Two different strategies for triggering are typically used in Cherenkov telescopes based on fast digitizers [6]. A majority trigger architecture consists of the discrimination of the signals coming from pixels, in order to count the number of activated pixels in a region. After discrimination, the signal processing can be analogue or digital. The sum trigger strategy is to sum the pulses from the pixels in a region and compare them with a threshold defined for the whole region. In a multilevel trigger architecture, these two strategies can both be implemented in the lower level (L0) trigger, while the higher level (L1) trigger circuit remains the same.

Our trigger system is designed for cameras made up of 7-pixel modules. One module consists of PMTs, front-end circuits, digitization and readout electronics, and trigger decision and distribution electronics.

### 2.2 Architecture of the trigger

The signal of every pixel is amplified and feeds a fast readout digitizer like the NECTAr [7] or DRAGON [8] readout boards. A two-level trigger scheme has been developed that is compatible with the electrical and mechanical architecture of some of the telescopes' cameras proposed for CTA. This scheme is shown in Figure 1.

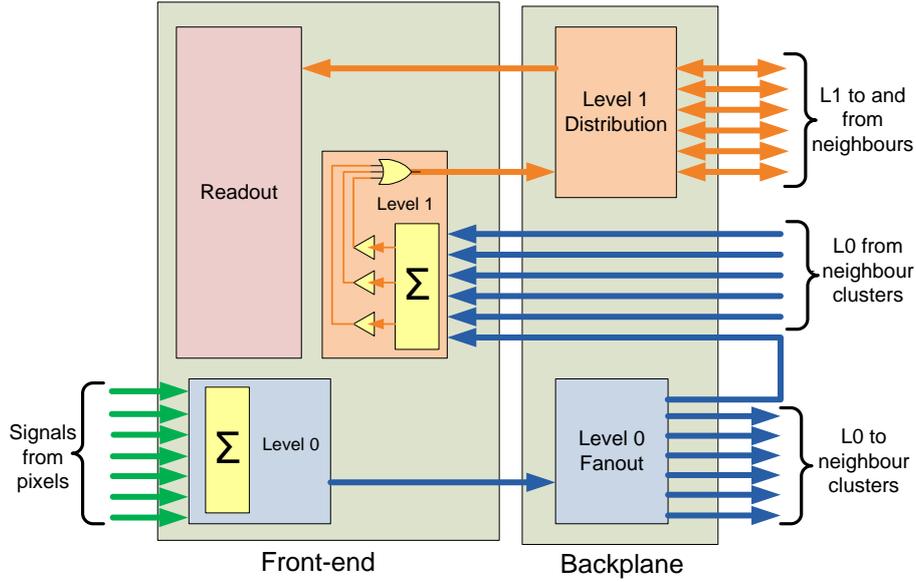

**Figure 1. Block diagram of the two-level analogue trigger.**

The first level, which we call L0, is module-based and combines the signals of the pixels in each module. In this level, the trigger implements the majority and sum trigger strategies, as shown in Figure 2. The Sum trigger analogically adds the signals from all pixels in the module. Before adding the signals from the individual pixels, each of them goes through attenuator and clipping circuits (both slow-control adjustable). The former allows all pixel gains to be equalized with a precision better than 5%. The later cuts the signals greater than a given value, which limits the influence of after-pulses from the photosensors. The second level, called L1, is also implemented in each module and combines the L0 signals of neighbouring modules from specific trigger regions with the local L0 signal. The L1 will generate a digital trigger pulse, which is distributed to the digitizers of all modules in the camera and is the final camera trigger. A detailed description of the ASIC implementing the L0 subsystem is presented in Section 3.

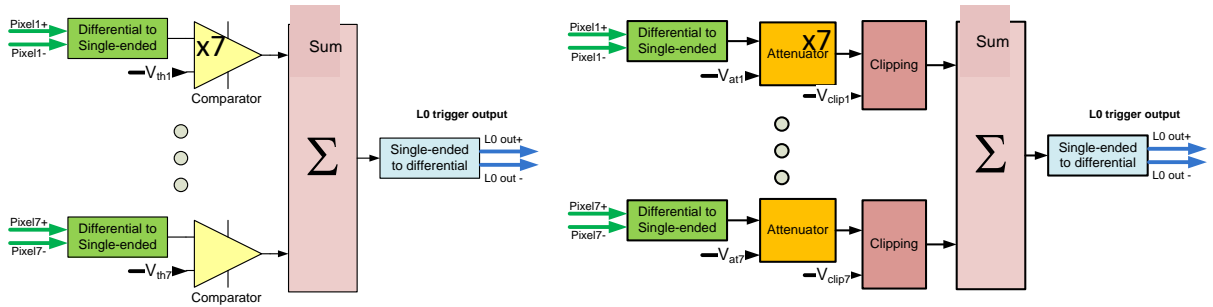

**Figure 2. Simplified block diagrams of the L0 trigger illustrating the majority (left) and sum (right) triggering schemes.**

## 3. ASIC design

Our mixed signal ASIC is designed to implement several low-level trigger functionalities. A block diagram is shown in Figure 3.

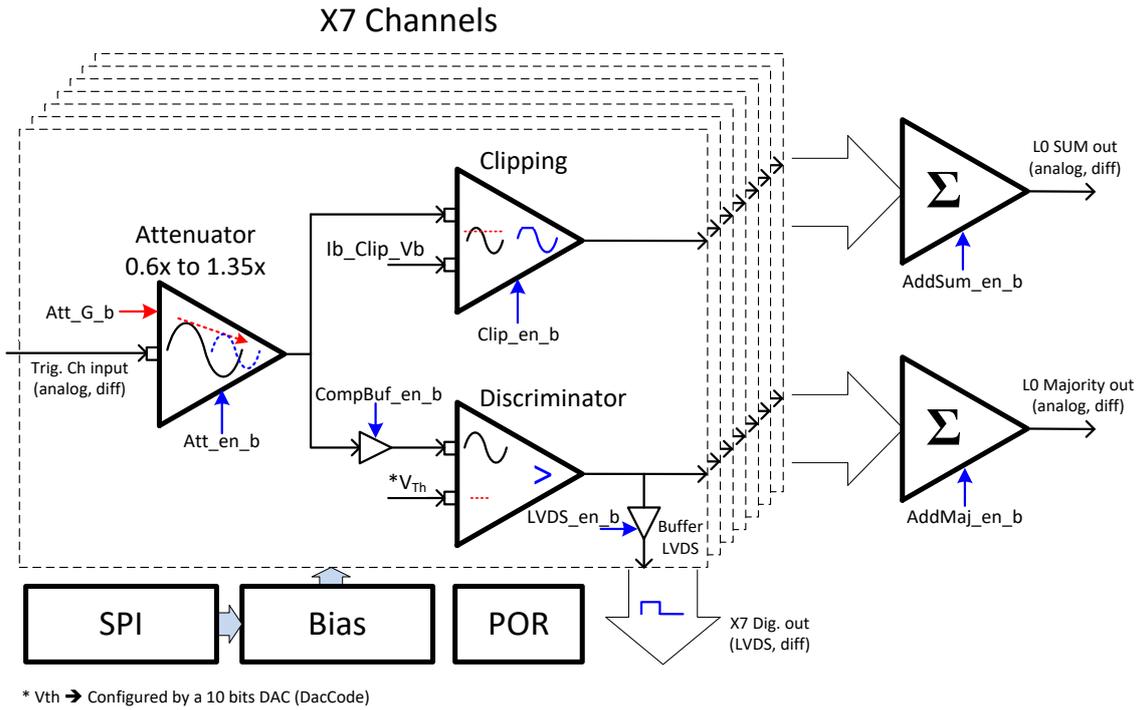

**Figure 3. Block diagram of the L0 ASIC.**

Both majority and sum trigger analogue schemes are implemented. The output of each majority discriminator is externally accessible, so a simple majority decision or a high-level trigger scheme can be implemented in a programmable logic device (typically an FPGA or CPLD).

    The requirements of the ASIC are summarized in Table 1. The L0 trigger receives the signal from each individual PMT after signal conditioning by a preamplifier with a bandwidth larger than 350 MHz (see [10] and [11]). The L0 input signal will have an amplitude of 20 mV per phe (photoelectron) with a signal to noise ratio (S/N) larger than four. The L0 can accommodate up to 20 phe per channel. A Gaussian with 2.4 ns FWHM is a good approximation of the signal shape.

| **Dynamic range (input)** | > 20 phe per channel |
|---|---|
| **Bandwidth** | > 500 MHz |
| **Output noise (sum)** | < 2 mV RMS (with unconnected inputs) |
| **Output noise (majority)** | < 0.2 phe (with unconnected inputs) |
| **Comparator DAC resolution** | > 8 bits |
| **Comparator efficiency** | > 95% |
| **Comparator purity** | > 99.99% |
| **Linearity error** | < ± 5% or 0.25 phe |
| **Gain adjustment** | Reduce 25% RMS amplitude dispersion to 5% RMS |
| **Clipping DAC resolution** | > 8 bits |
| **Power consumption** | < 150 mW/ch |

**Table 1. L0 trigger ASIC specifications.**

For the majority trigger, the input signal of each pixel passes through a comparator with a programmable threshold with an 8-bit range, and is then added to the other 6 pixels that are part of the module. The comparator should be able to react fast enough to the rising edge of the input pulse that it triggers on the signal provided by one phe with an efficiency larger than 95%, while keeping the probability to trigger on noise below 0.01%. The output of the comparator should be a square pulse with a width equal (within ±15%) to the time the input signal was above the discriminator threshold. The linearity error of the comparator, defined as (Mean-Fit)/Fit, should be at most ±5% or ±0.25 phe, whichever is larger.

For the sum trigger, there should be a gain adjustment, which does not need to be continuous. The amplitudes of the signals coming from the PMTs are expected to have an RMS deviation of around 25%. The gain adjustment should be able to reduce the amplitude spread to 5% RMS. The adjusted signal should go through a clipping block controlled by a DAC with a resolution of 8-bits or more.

Although both schemes are implemented, it is possible to power down the one that is not currently being used.

### 3.1 Basic differential stage

Several building blocks of the ASIC are based on different versions of the open loop differential stage depicted in Figure 4. Those blocks comprise the input attenuators, the clipping stage, and a transconductance stage in the adders, shown in Figure 3. An open loop fully differential architecture has been chosen to fulfil high-speed and low-power requirements. The main drawback of such an architecture is a moderate linearity compared with a closed loop solution. However, our linearity requirements are moderate, and closed loop (OTA/OPA based) solutions often suffer from linearity limitations due to slew rate problems.

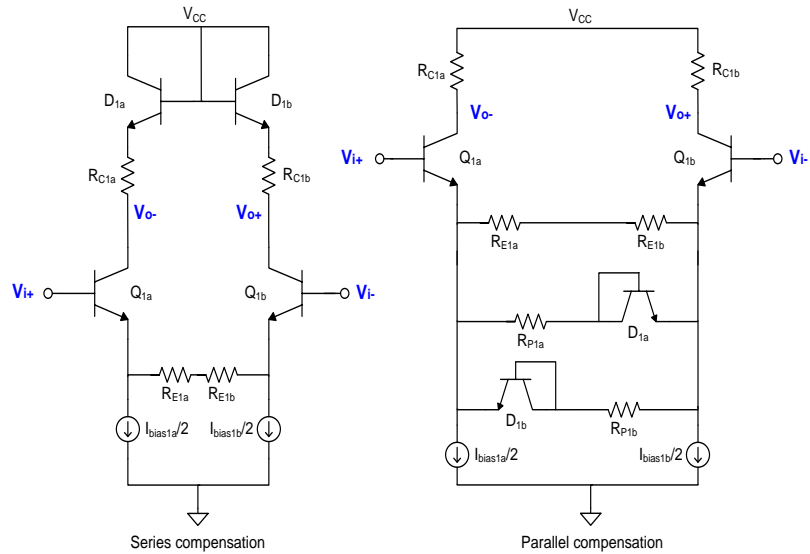

**Figure 4. Open loop differential stage. Series (left) and parallel (right) linearity compensation.**

The non-linearity of a differential pair is due to the variation of the base-emitter voltage of the two transistors. It is often corrected by the addition of one diode-connected transistor (gain of 1), as shown in Figure 4 ($D_{1a}$ and $D_{1b}$), or two diode-connected transistors (gain of 2) in the collector branches to obtain the same voltage drop in the emitter and collector. However, the consequence of this "serial correction" is lower voltage headroom and that the compensation is dependent on the gain. This is problematic for low-voltage operation and for circuits with variable gain, as is the case for this ASIC. In [12], a parallel correction scheme is presented. In this parallel correction, the loss of gain is compensated for by decreasing the emitter generation in the opposite branch, as shown in Figure 4. The new linearity compensation scheme permits compensation of different gains and does not incur any penalty in voltage headroom.

### 3.2 Discriminators

The discriminator block for the majority trigger mode is based on a differential discriminator and a 10-bit digital to analogue converter (DAC) designed for the SCOTT chip ([13] and [14]). In order to maximize speed, the discriminator utilizes a multistage architecture followed by a digital restorer and a latch, as depicted in Figure 5.

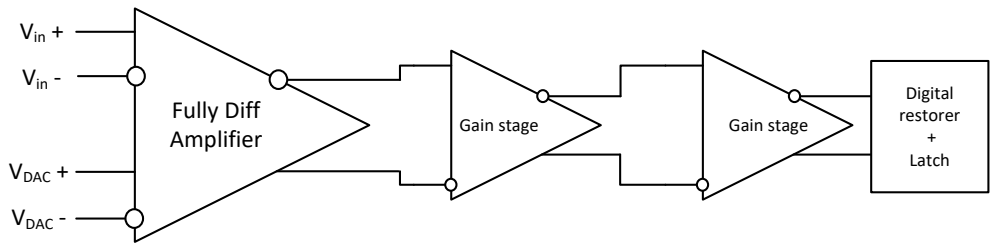

**Figure 5. Architecture of the discriminator.**

The DACs are differential and implement a resistor string architecture [15] to guarantee their monotonicity.

### 3.3 Adders

The fully differential adders for both the sum and majority trigger are designed as shown in Figure 6. Each adder consists of seven open loop transconductance stages (one per input) and a closed loop transimpedance amplifier based on a fully differential operational amplifier. The transconductance stage is based on a degenerated differential pair with the parallel scheme for linearity compensation presented in Figure 4.

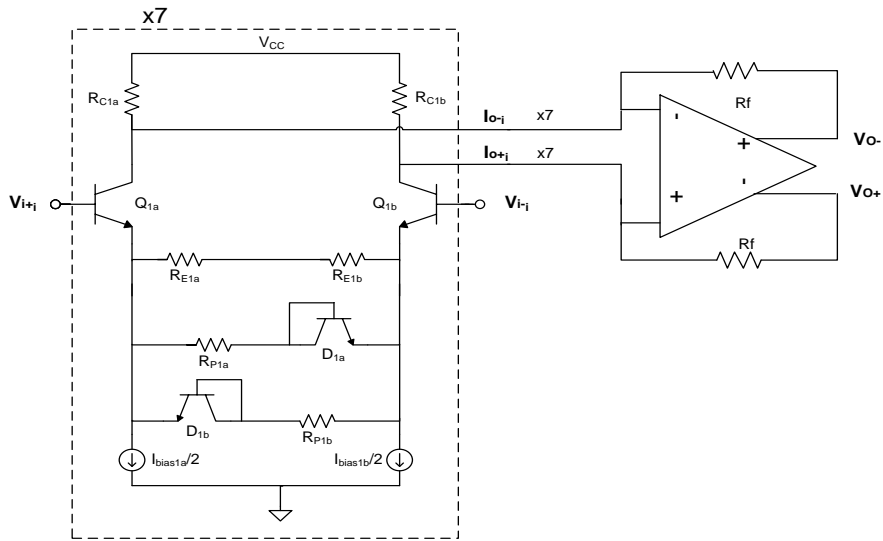

**Figure 6. Schematic of the adder block.**

The closed loop transimpedance amplifier in Figure 6 is based on a high gain-bandwidth product (GBP) operational amplifier. It is also used in other ASICs proposed for use in CTA [11]. A simplified schematic of a fully differential operational amplifier is shown in Figure 7 (left). It is a folded cascode amplifier, with a second Miller stage and a class AB output stage. Miller compensation is used with a nulling resistor. The input pair is degenerated to improve slew rate. Fast and accurate common-mode feedback is continuously provided by an error amplifier.

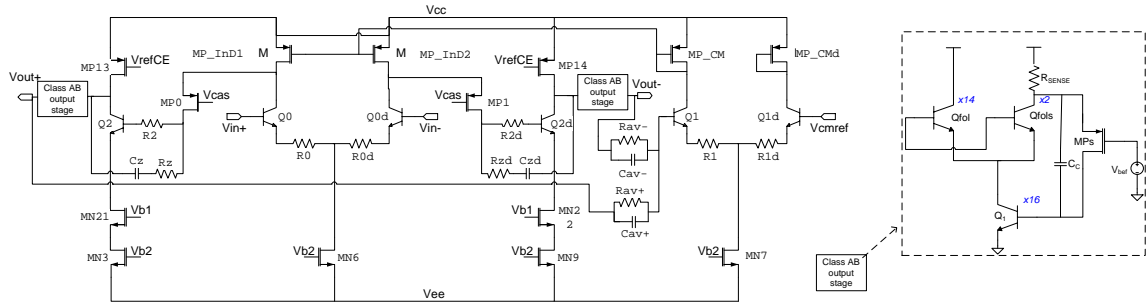

**Figure 7. Simplified schematic of the fully differential operational amplifier (left) and of the class AB output stage (right).**

The class AB output stage is shown in Figure 7 (right). The push-pull operation is based on a fast local feedback loop, where the GBP exceeds 2 GHz with a phase margin (PM) exceeding 60 deg. This stage can provide more than 20 mA peak current, with a quiescent current of 5 mA. This allows driving low load impedances with AC coupling, i.e., the cable or transmission line impedance connecting the L0 ASIC to the trigger backplane, as shown in Figure 1.

### 3.4 A 7-channel ASIC

The ASIC is implemented in a 0.35 µm BiCMOS technology with SiGe NPN HBTs. The use of heterojunction bipolar transistors (HBTs) is crucial for several reasons:

- The bandwidth (BW) of the signal processing in the ASIC must be higher than 500 MHz. This means that the bandwidth of each stage should be significantly higher, 700 MHz or more. Therefore, the use of high unity gain frequency $f_t$ active devices is crucial.

- The differential pair stages depicted in Figures 4 and 6 rely on emitter degeneration to achieve the required linearity. Bipolar devices provide a very good transconductance over bias current ratio.

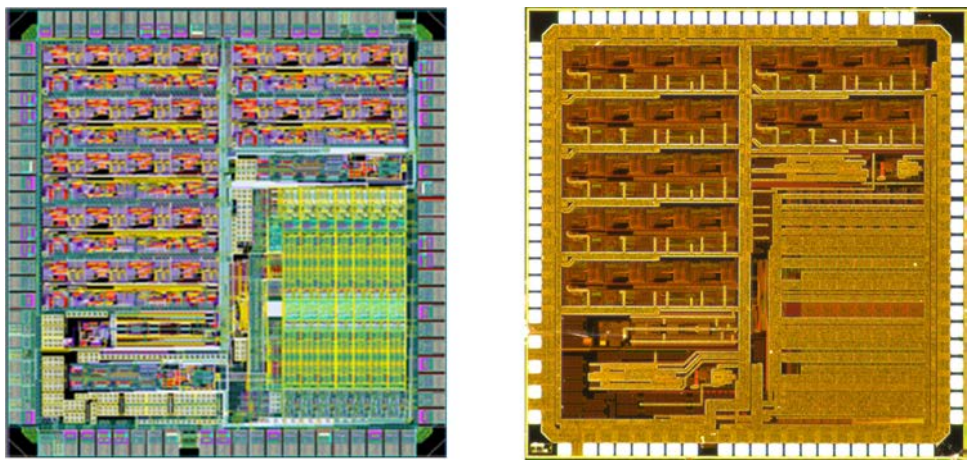

**Figure 8. Layout (left) and microphotograph (right) of the ASIC.**

The layout and microphotograph of the ASIC are shown in Figure 8. The area is about 12 mm² and it is packaged in a QFN 56 package.

Three different simultaneous operation modes are possible: sum trigger, majority trigger, and a digital trigger with the seven discriminated LVDS outputs. The subsystem corresponding to each mode can be set in "power down" mode independently for each individual channel. The chip can be configured and the operating parameters (thresholds, attenuation factors, bias currents, etc.) set by means of a Serial Peripheral Interface (SPI) bus. The power consumption depends on the ASIC configuration, but consumption for either sum or majority trigger operation is typically about 90 mW per channel.

## 4. Results

The L0 ASIC has been integrated in the NECTAr and DRAGON front end (FE) boards [16]. A typical L0 ASIC sum trigger output for a 2 phe signal using the DRAGON FE board is shown in Figure 9.

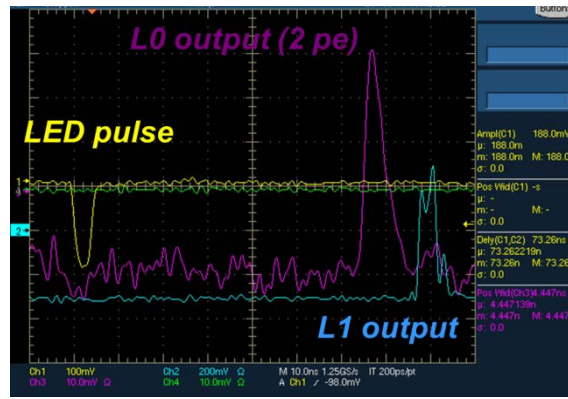

**Figure 9. Typical L0 ASIC sum trigger output for a 2 phe signal using DRAGON FE.**

The response of the sum trigger system for different clipping levels is depicted in Figure 10. The system is linear until a saturation or clipping level is reached. The clipping level is a configurable parameter.

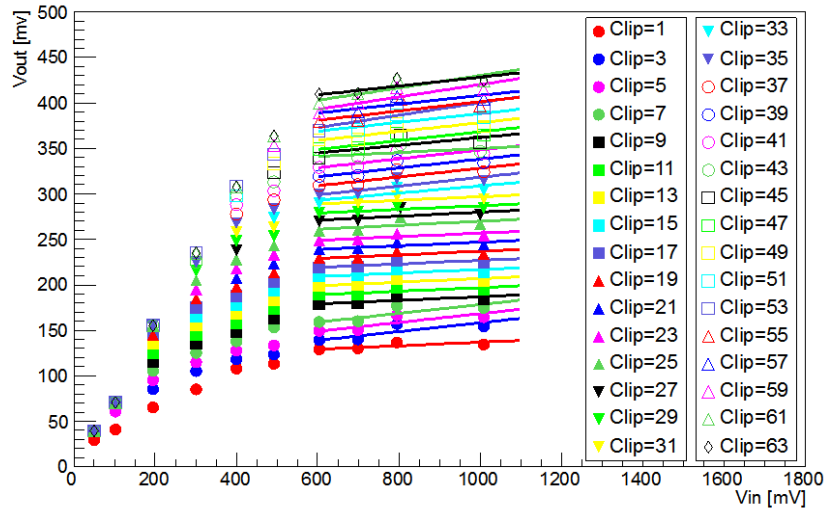

**Figure 10. Sum trigger transfer function for different fine clipping levels. Fine clipping is encoded in a 6-bit word, so the fine clipping range is between 0 and 63.**

The clipping level can be adjusted from 30 to 450 mV, as shown in Figure 11. Three different ranges can be selected by a coarse control (2 bits). A fine control (6 bits) sets the precise level of clipping for each channel.

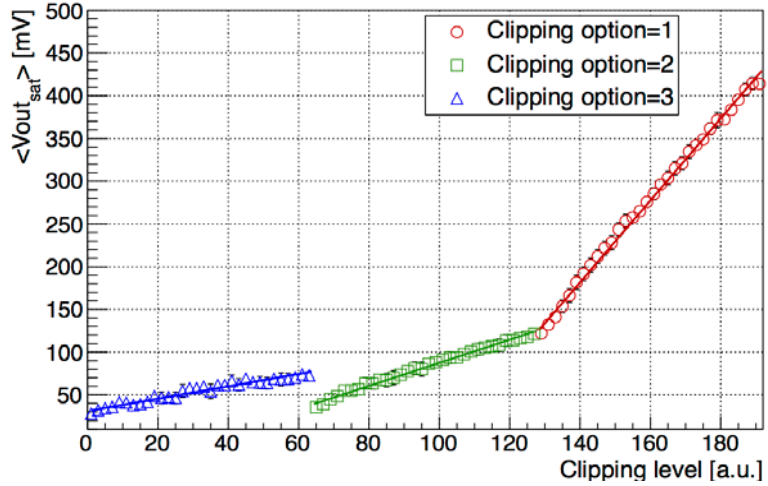

**Figure 11. Output voltage saturation level plotted as a function of clipping level (course and fine). Coarse control allows setting 3 different clipping regions (labelled clipping options in the plot).**

The performance of the sum trigger adder is illustrated in Figure 12. For a given attenuation and clipping, we measured the output of the adder. We used an input signal of 50 mV and a high clipping value in order to not clip the signals at the output. We measured the output of the adder for all the possible combinations of channels added. We histogrammed the measured outputs of the adder (divided by the number of channels added) for each number of channels added. Since the inputs for each channel are identical, the ratio of the adder's output voltage to the number of channels added should be the same in all cases. The histograms are fit with Gaussian functions whose standard deviations are at the level of 1-2% of the mean.

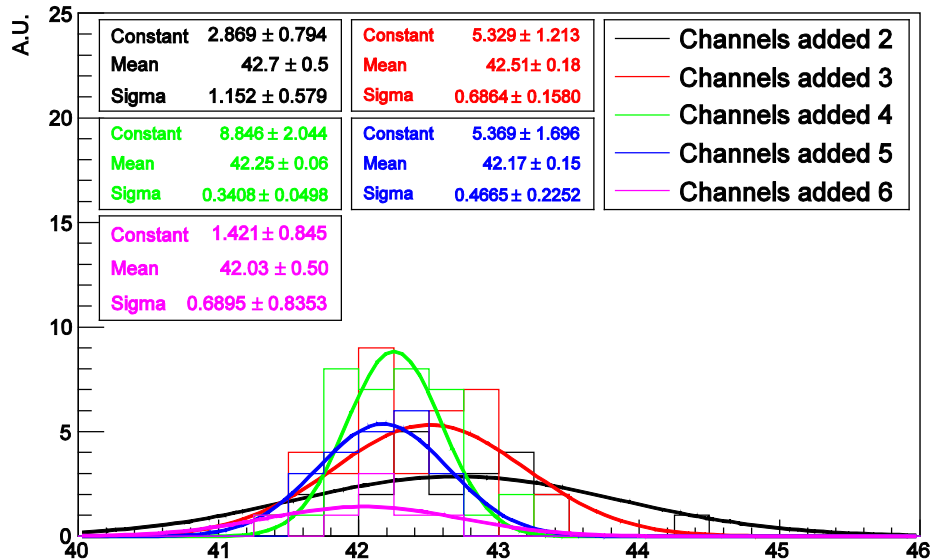

**Figure 12. Output of the sum adder divided by the number of channels added for an input signal of 50 mV. Histograms for the different number of channels added are shown in different colors (2 channels in black; 3 channels in red; 4 channels in green; 5 channels in blue; 6 channels in magenta) and fit with a Gaussian, represented by a line of the same color as the histogram. The results of the fit are shown in the plot with the same color as the fit.**

The majority trigger dynamic range, and thus discriminator linearity, is shown in Figure 13. It is important to verify the linearity of the discriminators so that the calibration of the discriminator thresholds provides an accurate measurement of the telescope energy threshold. The discriminator transition level as a function of input signal amplitude is obtained by a discriminator threshold scan. We define the discriminator threshold (DT) for a given input signal with a given attenuation as the discriminator voltage at which the output signal goes to zero. For the majority option, it is important that the DT increases linearly with the input voltage. This linearity was determined by measuring the DT of all the channels for a given input and attenuation. We plotted the mean of all the channels. We fit the mean DT for several inputs and found that the mean DT as a function of the input is linear within ~5% for any input. An example of the mean DT of one chip as a function of the input voltage, for a given attenuation, is shown in Figure 13.

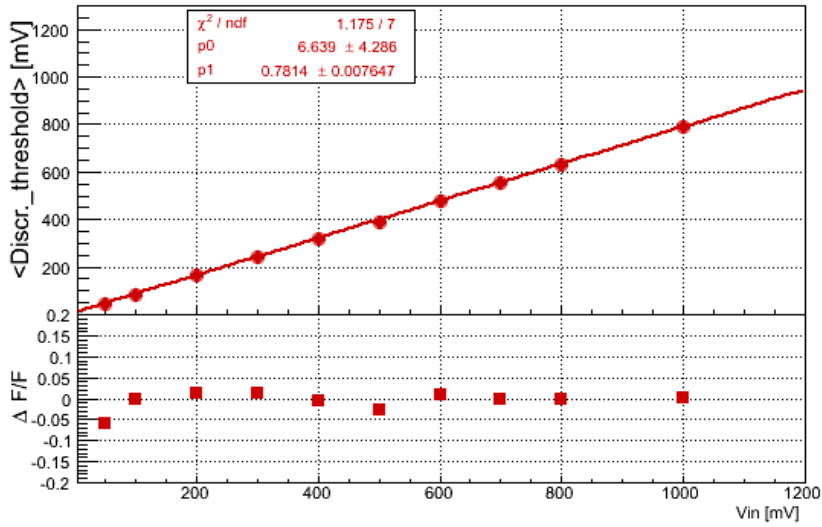

**Figure 13. Discriminator transition for different input signal amplitudes. Linear fit (top) and residuals (bottom) are shown.**

The noise of the majority chain can be inferred from the curves resulting from the threshold scans. The trigger rate as function of the threshold value for the seven channels of an ASIC is depicted in Figure 14. The noise is about 1 mV RMS, i.e. well below 0.2 phe.

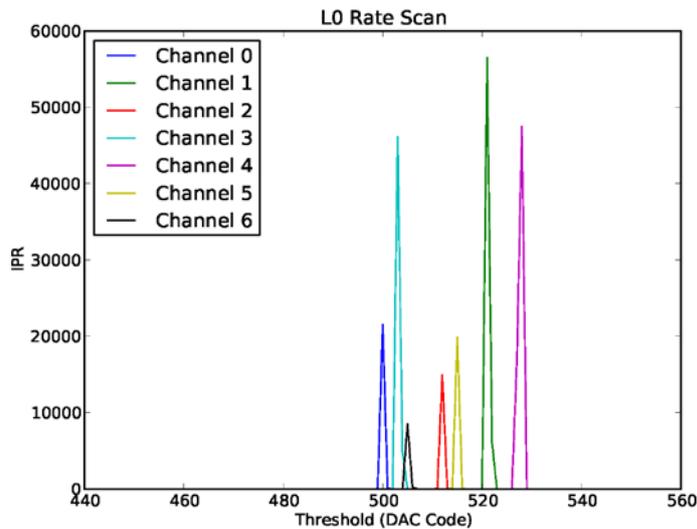

**Figure 14. Trigger Individual Pixel Rate (IPR) as function of threshold for the 7 ch of an L0 ASIC.**

The majority trigger timing characteristics are shown in Figure 15. Relative delay (from input to output) variation and output pulse width are measured as a function of input signal amplitude. The input pulse FWHM is about 3 ns. The threshold is set to about 6 phe, and the approximate calibration is 20 mV/phe at the input of the L0 ASIC. The discriminator propagation delay and time walk are well below 2 ns.

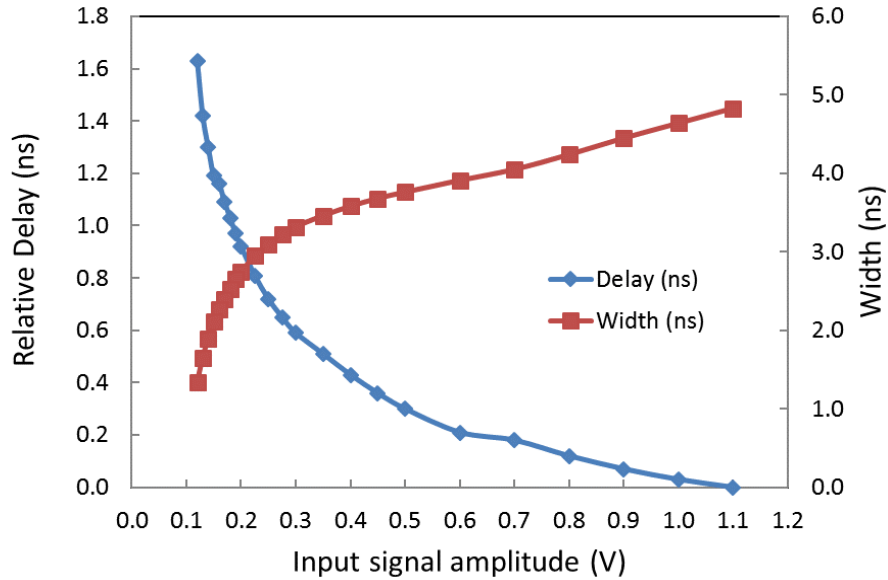

**Figure 15.** Majority trigger timing characteristics. Relative delay variation (blue) and output pulse width (red) are plotted as a function of input signal amplitude (input FWHM is about 3 ns). The threshold is set to about 6 phe, with 20 mV/phe at the input of the L0 ASIC.

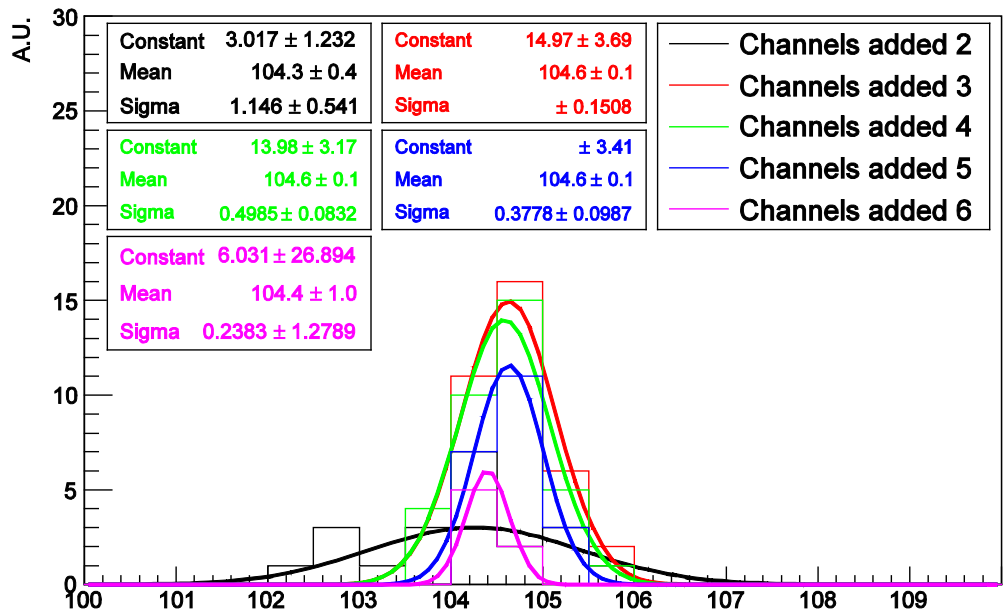

**Figure 16.** Output of the majority adder divided by the number of channels added. Histograms for the different number of channels added are shown in different colors (2 channels in black; 3 channels in red; 4 channels in green; 5 channels in blue; 6 channels in magenta) and fit with a Gaussian, represented by a line of the same color as the histogram. The results of the fit are shown in the plot with the same color as the fit.

The performance of the majority adder is illustrated in Figure 16. In this case, the discriminator outputs are added. For a given attenuation, we measured the addition of all the possible combinations of channels. The final output was divided by the number of channels added to fill different histograms, depending on the number of channels added. The histograms were fitted with Gaussian functions with standard deviations of < 1% of the mean normalized output voltage.

## 5. Conclusions

We have presented a versatile and reconfigurable ASIC for triggering Cherenkov telescope cameras. The ASIC implements two different analogue trigger schemes: the sum trigger and the majority trigger. Furthermore, the discriminator outputs are available as independent LVDS signals, allowing for the implementation of complementary digital trigger schemes, such as the FPGA-based digital trigger presented in [16].

Unused blocks in a given operation mode can be individually set in standby mode, thus optimizing the power consumption, which is a factor of five smaller than equivalent solutions implemented with commercial off-the-shelf components [6].

A combination of open and closed loop design techniques is used to achieve high speed and minimize power consumption, while fulfilling linearity and noise requirements. Fully differential design is used to increase power supply noise rejection and to increase dynamic range.

## Acknowledgments

Funding for this work was partially provided by the Spanish MINECO and UE FEDER under projects: FPA2010-22056-C06-02, FPA2010-22056-C06-01, FPA2015-69210-C6-2-R, and MDM-2014-0369 of ICCUB; SEV-2012-0249 of IFAE; FPA2010-22056-C06-06, FPA2013-48381-C06-2-P, and FPA2014-55819-C4-3-P of UCM; and finally FPA2010-22056-C06-03, FPA2013-48381-C6-3-P, and FPA2014-55819-C4-2-P of CIEMAT.
This paper has gone through internal review by the CTA Consortium.
We would like to thank the reviewers and committee (J.F. Glicenstein, M. Daniel, R. Paoletti and J. Rico) for their useful suggestions.

## References

[1] CTA Project, *Cherenkov Telescope Array*, [Online] Available: http://www.cta-observatory.org/

[2] MAGIC Collaboration *Major Atmospheric Gamma-ray Imaging Cherenkov Telescope* [Online] Available: http://http://wwwmagic.mppmu.mpg.de/index.html

[3] HESS Observatory *High Energy Stereoscopic System* [Online] Available: http://www.mpi-hd.mpg.de/hfm/HESS

[4] VERITAS Observatory *Very Energetic Radiation Imaging Telescope Array System* [Online] Available http://veritas.sao.arizona.edu/

[5] M. Rissi, N. Otte, T. Schweizer and M. Shayduk, "A New Sum Trigger to Provide a Lower Energy Threshold for the MAGIC Telescope," in IEEE Transactions on Nuclear Science, vol. 56, no. 6, pp. 3840-3843, Dec. 2009. doi: 10.1109/TNS.2009.2030802

[6] M. Barcelo, J.A. Barrio, O. Blanch Bigas, J. Boix, C. Delgado, D, Herranz, R, Lopez-Coto, G. Martinez, L.A. Tejedor *An Analog Trigger System for Atmospheric Cherenkov Telescopes*, in Nuclear

Science, IEEE Transactions on , vol.60, no.3, pp.2367-2375, June 2013, doi: 10.1109/TNS.2013.2257852

[7] J. F. Glicenstein, et al, *The NECTAr project: a New Electronics design for Cherenkov Telescope Arrays Arrays*, Proc. 32nd Int. Cosmic Ray Conf., Beijing, China, 2011.

[8] H. Kubo, R. Paoletti, et al, *Development of the Readout System for CTA Using the DRS4 Waveform Digitizing Chip*, Proc. 32nd Int. Cosmic Ray Conf., Beijing, China, 2011.

[9] Luis A. Tejedor, José I. Alonso, Juan A. Barrio, José L. Lemus, Carlos Delgado *An Analog Level 1 Trigger Prototype fot CTA*, in Nuclear Science, IEEE Transactions on , vol.60, no.1, pp.251-258, Feb. 2013, doi: 10.1109/TNS.2012.2226246

[10] D. Gascon, A. Sanuy, E. Delagnes, J. Sieiro, F. Feinstein, J.F. Glicenstein, P. Nayman, M. Ribo, F. Toussenel, J.P. Tavernet, P. Vincent, S. Vorobiov. *Wideband pulse amplifier with 8 Ghz GBW product in 0.35 µm CMOS technology for the integrated camera of the Cherenkov Telescope Array*, Journal of Instrumentation, Volume 5, December 2010, http://stacks.iop.org/1748-0221/5/i=12/a=C12034.

[11] A. Sanuy, D. Gascon, J.M. Paredes, L. Garrido, M. Ribó, J. Sieiro, *Wideband (500 Mhz) 16 bit dynamic range current mode PreAmplifier for the CTA cameras*, in Nuclear Science Symposium and Medical Imaging Conference (NSS/MIC), 2011 IEEE , vol., no., pp.750-757, 23-29 Oct. 2011, doi: 10.1109/NSSMIC.2011.6154096.

[12] J. Lecoq, G. Bohner, R. Cornat, P. Perret and C. Trouilleau, *The mixed analog/digital shaper of the LHCb preshower*, 7th Workshop on Electronics for LHC Experiments, 2001

[13] S. Ferry, F. Guilloux, S. Anvar, F. Chateau, E. Delagnes, V. Gautard, F. Louis, E. Monmarthe, H. Le Provost, S. Russo, J.-P. Schuller, Th. Stolarczyk, B. Vallage, E. Zonca, *Multi-time-over-threshold technique for photomultiplier signal processing: Description and characterization of the SCOTT ASIC*, Nuclear Inst. and Methods in Physics Research, A, Volume 695, 11 December 2012, Pages 52-60, ISSN 0168-9002, http://dx.doi.org/10.1016/j.nima.2011.10.004.

[14] S. Ferry, F. Guilloux, S. Anvar, F. Chateau, E. Delagnes, V. Gautard, F. Louis, E. Monmarthe, H. Le Provost, S. Russo, J-P. Schuller, Th. Stolarczyk, B. Vallage, E. Zonca, *SCOTT: A time and amplitude digitizer ASIC for PMT signal processing*, Nuclear Inst. and Methods in Physics Research, A, Volume 725, 11 October 2013, Pages 175-178, ISSN 0168-9002, http://dx.doi.org/10.1016/j.nima.2012.11.164.

[15] C.-C. Chen, et al., *A 10-bit folded multi-LSB decided resistor string digital to analog converter*, in: Proceedings of the International Symposium on ISPAC, 2006, pp. 123–126.

[16] H. Kubo et al., *Development of the photomultiplier tube readout system for the first Large-Sized Telescope of the Cherenkov Telescope Array*, In Proceedings of the 34th International Cosmic Ray Conference (ICRC2015), The Hague, The Netherlands, 2015.